\documentclass[12pt]{article}
\usepackage[top=1in, bottom=1in, left=1in, right=1in]{geometry}

\usepackage{./dsm5_header}

\setTitle{\OCCA: A unified approach to multi-threading languages}

\addAuthor{David S. Medina \setFN{\fnStar   }\footnote{David S. Medina, Computational and Applied Mathematics at Rice University (dsm5@rice.edu)}}
\addAuthor{Amik St-Cyr     \setFN{\fnDagger }\footnote{Amik St-Cyr, Computation and Modeling at Shell  (Amik.St-Cyr@shell.com)}}
\addAuthor{T. Warburton   \setFN{\fnDDagger}\footnote{Tim Warburton, Computational and Applied Mathematics at Rice University (timwar@rice.edu)}}

\begin{document}

\placeTitlePage

\setCode{occa,none,\scriptsize}

\begin{outline}
The inability to predict lasting languages and architectures led us to develop \occa, a C++ library focused on host-device interaction.
Using run-time compilation and macro expansions, the result is a novel single kernel language that expands to multiple threading languages.
Currently, \occa supports device kernel expansions for the OpenMP, OpenCL, and CUDA platforms.
Computational results using finite difference, spectral element and discontinuous Galerkin methods show \occa delivers portable high performance in different architectures and platforms.
\end{outline}

\section{Introduction}

Until 2001, the top supercomputers were built on a parallel architecture involving multiple single-core vector machines.
Multi-core processors then became the standard with co-processor acceleration as early as 2005 \cite{top500}.
Not knowing which architectures will stand the test of time places programmers in an awkward position.
For instance, IBM's cell architecture, known for as the graphics accelerator choice in the Playstation 3, was used in IBM's Roadrunner, the first petaflop-capable machine.
Although the cell architecture had high promise, it was put on hiatus less than a decade after development \cite{cellNews}.
Graphical processing units (GPUs) quickly dominated the co-processor competition with the only current competitor being Intel's Xeon Phi.
Although few supercomputers make use of the Xeon Phi as an accelerator, 4 out of the 6 systems listed in the top500 are currently the top 10 in performance, including the top supercomputer in the list \cite{top500}.

\subsection{GPU Programming}

Initially, instructions were passed from the CPU to the GPU with the use of programmable graphics shaders, mainly using OpenGL's shading language (GLSL), DirectX's high-level shader language (HLSL), or NVIDIA's Cg shading language.
As early as 2004, shaders began to be used for general purpose computations outside of graphics.
The Brook language was developed to reduce the complexity of using graphics shaders for general purpose computations \cite{buck2004brook}.
Brook was a predecessor to NVIDIA's CUDA, extending the C language to include GPU variable and function declarations to hide shader-programming complexity.
NVIDIA released CUDA in 2007 and extended the C++ language to allow GPU programming in a similar manner to the Brook language.
Adopting CUDA for development vendor-locks the implementation to NVIDIA GPUs for hardware acceleration \cite{bodin2009heterogeneous}.

The Khronos group, a consortium focused on creating open standards, released the OpenCL standard for programming heterogeneous platforms including CPUs, GPUs, Intel's Xeon Phi and field-programmable gate arrays (FPGAs).
A year after CUDA's release, NVIDIA and Advanced Micro Devices (AMD) gave their first functional OpenCL demo in Siggraph Asia 2008 \cite{munshi2008opencl}.
Implementations of the OpenCL standard, however, provide a C library API rather than a language extension.

\subsection{Compiler Constructs}

New standards, languages and extensions often accompany the arrival of new architectures, such as OpenMP for multi-core processors and both CUDA and OpenCL for GPUs.
Refactoring legacy codes to suit new architectures with minimal adjustments has been a recurring challenge, inspiring standards such as OpenMP and OpenACC\cite{dolbeau2007hmpp}.

OpenMP and OpenACC standards focus on giving hints to compilers to perform specific optimizations for parallelizing code regions with the use of pragmas (optional directives).
Using supported compilers, code regions can be parallelized to use thread-based parallelization (OpenMP) or GPU acceleration (OpenACC) with minimal code modification.
Although compiler-aided parallelization is achieved with pragma insertions, there are several obstructions to automating parallelization.
The use of OpenMP requires the use of only thread-safe functions and knowledge about memory residency for variables to prevent read-write conflicts.
Similarly, OpenACC has memory residence ambiguity from pointer aliasing, often requiring memory transfers between the host and device before and after critical regions.

subsection{Language Interface}

Apart from compiler-based solutions for refactoring, we introduce different approaches to platform flexibility on GPU-accelerated projects.
As previously mentioned, CUDA works only on NVIDIA cards; and as a result, there are many projects are aimed on automating translations between CUDA and OpenCL for device-portability.
Aside from using compiler pragmas for code generation, we discuss projects that create translation layers between languages using code translation or a unifying API.
Swan \cite{harvey2011swan} and CU2CL \cite{martinez2011cu2cl} were two of the first of few projects to attempt a conversion between CUDA and OpenCL, both using distinct source-to-source methods.

Swan is a perl-based library that uses regular expressions to translate CUDA kernels to OpenCL kernels.
From AMD's ``Porting to OpenCL'' \cite{amdporting2013} website, it is clear that most of the CUDA to OpenCL kernel translations can be done with string replacements.
Swan also provided an API that shadowed CUDA and OpenCL runtime API to hide the difference in API between the two languages.

On the other hand, CU2CL used clang's preprocessor, abstract syntax trees and framework to facilitate CUDA to OpenCL code translation \cite{martinez2011cu2cl}.
The novelty in CU2CL is the ability to translate both kernels and source code, while the approach by Swan introduces an API layer to the source code.

GPU Ocelot is another project towards programming heterogeneous platforms.
Led by a group in Georgia Tech's Computer Architecture and Systems Laboratory (CASL), Ocelot is capable of transforming PTX (NVIDIA's intermediate parallel assembly language) to low level assembly languages in different platforms.
Targeted platforms include the CPU architecture through emulated kernels and AMD GPUs through a transformation to AMD's CAL assembly language \cite{diamos2010ocelot}.

So far, we discussed projects aimed on platform portability; we now introduce some portable libraries focused on mathematical functionalities rather than general purpose computations.
ViennaCL \cite{rupp2010viennacl} is an example of a library with linear algebra routines on multiple platforms.
ViennaCL is able to switch enable OpenMP, OpenCL and CUDA for parallel computations at compile-time and choose the back-end platform for executing kernels.
PyFR \cite{witherden2013pyfr} is a python library focused on numerical solver implementations that is able to use either OpenMP, OpenCL and CUDA for advection-diffusion type equations.
However, the kernel generations available in PyFR are fixed to their set of numerical routines available, rather than offering a generic API to code for the different platforms.

\subsection{\OCCA Overview}

The different projects mentioned have focused on creating some mapping between two or more programming languages for these accelerators, or switching between multiple platforms for computation purposes.
We present \occa, a library including an API that abstracts back-ends and kernel languages from OpenMP, OpenCL and CUDA.
Although \occa unites different threading platform back-ends, the main contributions is the abstraction of the kernel language.
Using a macro-based approach, an \occa kernel can be expanded at runtime to suit OpenMP with dynamic pragma insertions or a device kernel using either OpenCL or CUDA and be used to streamline the incorporation of future languages with ease.

The host API and kernel language in \occa were developed to maintain portability and performance together with platform-choice flexibility.
First, a thorough description on the host front-end in \occa will be given in \refSec{sec:host}.
Following is an explanation in \refSec{sec:device} on how the kernel formatting and language unwraps to OpenMP, OpenCL and CUDA.

Throughout this paper we use OpenCL notation but include CUDA's complement inside double brackets such as (OpenCL \opt{CUDA}) to avoid focusing on one terminology.
After \occa has been explained, we give in \refSec{sec:methods} different numerical method implementations utilizing \occa.
For further reference, an overview of \occa keywords is provided in the \hyperref[sec:appendix]{appendix}.

\section{\OCCA Host API}
\label{sec:host}

\begin{center}\begin{figure}[h!]
\centering
\includegraphics[width=10cm]{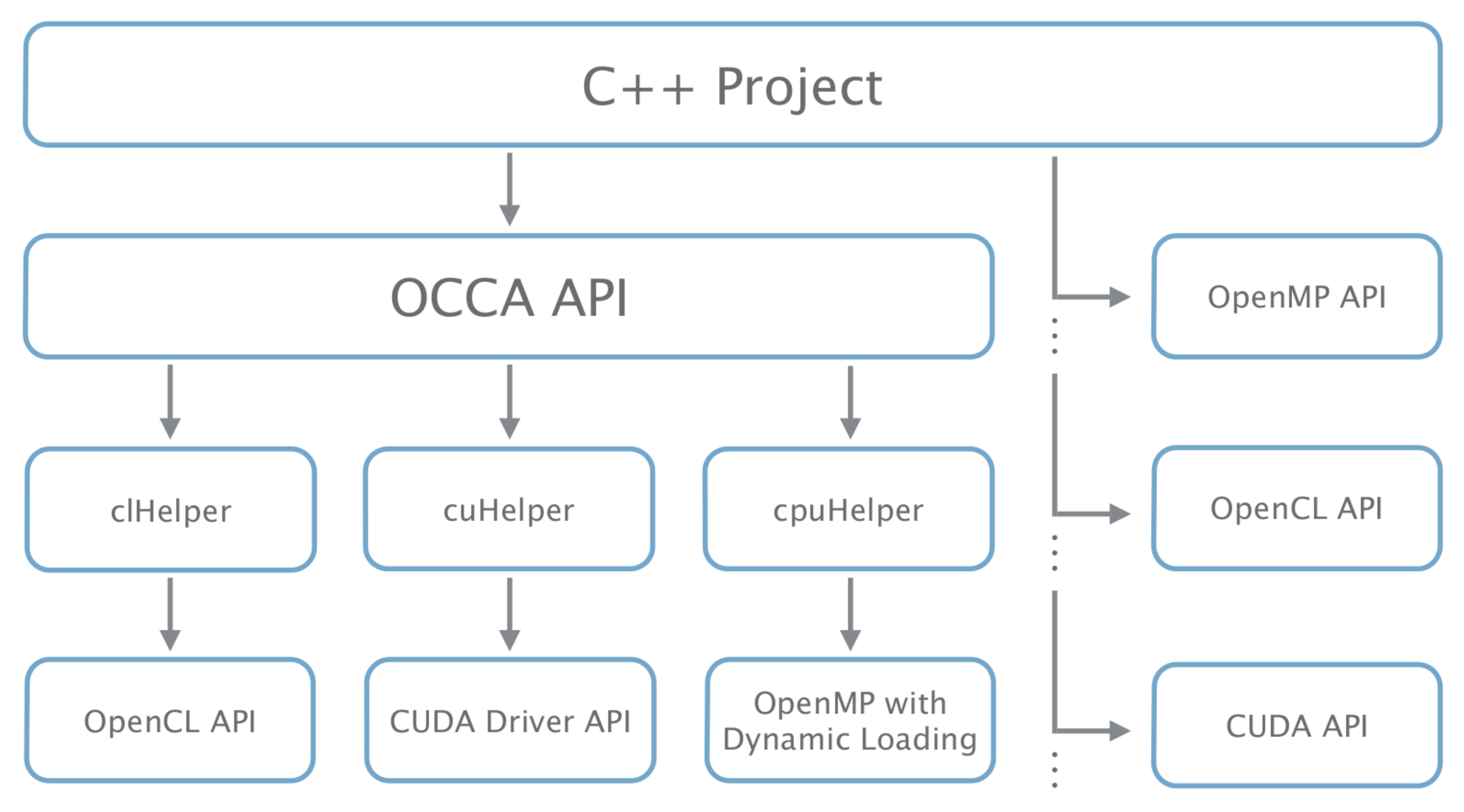}
\caption{\OCCA wraps different language APIs and is non-conflicting with external libraries in either platform}
\label{fig:flow}
\end{figure}\end{center}

Aside from language-based libraries from OpenMP, OpenCL or CUDA, the \occa host API is a stand-alone library.
The independence allows \occa to be combined with other libraries without conflict, as shown in \refFig{fig:flow}.

\vspace{4mm}
\vspace{-13mm}

Unlike many of the projects and codes already mentioned, the focus of \occa is to provide tools for multi-platform computing, rather than code transformation.
The following sections focus on the three key components that influenced the \occa host API development: the platform device, device memory and device kernels.

\subsection{\OCCA device}

An \occa device acts as a layer of abstraction between the \occa API and the API from supported languages.
Due to the just-in-time code generation discussed in \refSec{sec:device}, the platform target can be chosen at run-time.
Enabled platforms are managed at compile-time in the case of unsupported platforms on the compiled architecture.

An \occa device generates a self-contained context and command queue \opt{stream} from a chosen device, being a socketed processor, GPU or other OpenCL supported devices such as a Xeon Phi or an FPGA.
Asynchronous computations with multiple contexts can be achieved using multiple \occa devices.
The device's main purpose is to allocate memory and compile kernels for the chosen device.

\subsection{\OCCA memory}

The \occa memory class abstracts the different device memory handles and provide some useful information such as device array sizes.
Although memory handling in \occa facilitates host-device communication, the management of reading and writing between host and device, for performance reasons, is still left to the programmer.
Having a class dedicated for device memory also allows the \occa kernel class to differentiate and communicate between distinct memory types.

\subsection{\OCCA kernel}

The \occa kernel class unites device function handles with a single interface, whether for a function pointer (OpenMP), cl\_kernel (OpenCL), or cuFunction (CUDA).
When using the OpenCL and CUDA kernel handles, passing the arguments through their respective API is simple, but there are discrepancies when comparing to the OpenMP wrapper.
For example, OpenCL and CUDA kernels work-items \opt{threads} have access to work-group \opt{block} and work-item counts implicitly.
However, C++ functions only have access to the function scope and global namespace, requiring the work-group and work-item counts to be passed as macros or as an argument to the kernel.

\section{\OCCA Device API}
\label{sec:device}

The \occa device API heavily relies on macros to give the user a clear front-end for programming device kernels.
Defining grid and work-group size as a macro gives the kernel a fixed overall working size; changing the working size would require a kernel re-compilation.
Hence, we approached this problem with the addition of an extra argument to the device kernels:

\vspace{3mm}
\begin{lstlisting}[caption={An example of an \occa kernel header},label={lst:kernelHeader}]
occaKernel void kernelExample(occaKernelInfoArg,
                              occaPointer datafloat* arg1
                              occaPointer datafloat* arg2){}
\end{lstlisting}
\vspace{-2mm}

Macros were used to build the device API, letting the preprocessor translate the kernels to fit OpenMP, OpenCL, and CUDA standards.
This approach was chosen, as opposed to the source-to-source solution used in other projects, to create a simple and lightweight library adding the flexibility to support other languages.

The \texttt{occaKernelInfoArg} in \refLst{lst:kernelHeader} acts as a place-holder for external information needed in the kernel.
For now, \texttt{occaKernelInfoArg} is only used in OpenMP mode to pass the computational grid and work-group sizes.
The use of this place-holder will allow flexibility when enabling another platform or language.
A comprehensive collection of all \occa keywords and macro expansions can be found in the \hyperref[sec:appendix]{appendix}.

\subsection{Macro-based kernel generation}

Using macros to connect the multiple languages creates a precise translation that is intuitive to the programmer without code-transformation subtlety.
The \occa keywords span keywords used in OpenMP, OpenCL and CUDA to cover major functionalities, such as implicit for-loops, barriers, etc.

Large projects tend to maintain backwards compatibility, while adding new features through updates or patches.
Similarly, CUDA and OpenCL have kept a backwards-compatible model, a feature possible in \occa with the addition of new \occa keywords with the introduction of new macros.

\subsection{Handling multi-threading architectures}

The coding approach taken in both OpenCL and CUDA for coding multi-threading architectures is done at the granularity of a single work-item \opt{thread}.
Work-items are part of a work-group \opt{block} that are able to communicate efficiently using shared memory.

\setCode{occa,none,\scriptsize}
\vspace{3mm}
\begin{lstlisting}[caption={
  The code listing expands the implicit for-loops found in OpenCL kernels.
  Loop grouping (1) expands multi-dimensional work-groups \opt{blocks} and loop grouping (2) expands multi-dimensional work-items \opt{threads}.
},label={lst:clLoopExpansion}]
for(int bZ = 0; bZ < get_num_groups(2); ++bZ){        // (1)
  for(int bY = 0; bY < get_num_groups(1); ++bY){
    for(int bX = 0; bX < get_num_groups(0); ++bX){
      // Shared memory is initialized here

      for(int tZ = 0; tZ < get_local_size(2); ++tZ){  // (2)
        for(int tY = 0; tY < get_local_size(1); ++tY){
          for(int tX = 0; tX < get_local_size(0); ++tX){
             // Scope of the OpenCL kernels
             // Register memory is initialized here
      }}}
}}}
\end{lstlisting}
\vspace{-2mm}

\setCode{occa,none,\scriptsize}
\vspace{3mm}
\begin{lstlisting}[caption={
  Similar to \refLst{lst:clLoopExpansion}, the expansion of the implicit for-loops found in CUDA kernels is displayed.
  Loop grouping (1) expands multi-dimensional work-groups \opt{blocks} and loop grouping (2) expands multi-dimensional work-items \opt{threads}.
},label={lst:cuLoopExpansion}]
for(int bZ = 0; bZ < gridDim.z; ++bZ){        // (1)
  for(int bY = 0; bY < gridDim.y; ++bY){
    for(int bX = 0; bX < gridDim.x; ++bX){
      // Shared memory is initialized here

      for(int tZ = 0; tZ < blockDim.z; ++tZ){  // (2)
        for(int tY = 0; tY < blockDim.y; ++tY){
          for(int tX = 0; tX < blockDim.x; ++tX){
             // Scope of the CUDA kernels
             // Register memory is initialized here
      }}}
}}}
\end{lstlisting}
\vspace{-2mm}

The loop groupings \tf{(1)} and \tf{(2)} in \refLst{lst:clLoopExpansion} and \refLst{lst:cuLoopExpansion} are implicitly invoked in OpenCL and CUDA by obtaining block information (\tf{bX, bY, bZ}) and thread information (\tf{tX, tY, tZ}) through language-specific calls.
Instances from the loop grouping \tf{(1)} are generated independently (no order dependence is guaranteed in GPU programming) but each instance creates the threads seen in the grouping \tf{(2)}.
Because of this, we generalize the loop formations to suit both, GPU programming and the explicit thread-based architecture of OpenMP.
All \occa keywords related to loop expansions can be found in \refTab{tab:workSizes} of the appendix,

\setCode{occa,none,\scriptsize}

\vspace{3mm}
\begin{lstlisting}[caption={
  The \occa programming model mirrors GPU programming, where group loopings (1) and (2) refer to work-groups \opt{blocks} and work-items \opt{threads} respectively.
},label={lst:occaLoopExpansion}]
occaOuterFor2{       // Loop grouping (1)
  occaOuterFor1{
    occaOuterFor0{
      // Shared memory defined here

      occaInnerFor2{ // Loop grouping (2)
        occaInnerFor1{
          occaInnerFor0{
      }}}
}}}
\end{lstlisting}
\vspace{-2mm}

where each \tf{occaFor} macro is expanded with respect to the programming model chosen.
In the case of GPU languages, the macros are defined as empty macros, encompassing whole kernel inside nested scopes that are optimized away.
The kernel code expansion for the OpenMP mode is displayed in \refLst{lst:pragmaExpansion}.

\setCode{occa,none,\scriptsize}
\vspace{3mm}
\begin{lstlisting}[caption={
  Using a programming practice similar to GPU work-group \opt{block} and work-item \opt{thread} executions, our best choice to place the OpenMP pragma expansions were at the analogous inner-most work-group looping (1).
},label={lst:pragmaExpansion}]
//(1)
for(occaOuterId2 = 0; occaOuterId2 < occaOuterDim2; ++occaOuterId2){
  for(occaOuterId1 = 0; occaOuterId1 < occaOuterDim1; ++occaOuterId1){
    _Pragma("omp parallel for
            firstprivate(occaInnerId0, occaInnerId1, occaInnerId2,
                         occaDims0   , occaDims1   , occaDims2)")
    for(occaOuterId0 = 0; occaOuterId0 < occaOuterDim0; ++occaOuterId0){
      // Shared memory defined here

      // (2)
      for(occaInnerId2 = 0; occaInnerId2 < occaInnerDim2; ++occaInnerId2){
        for(occaInnerId1 = 0; occaInnerId1 < occaInnerDim1; ++occaInnerId1){
          for(occaInnerId0 = 0; occaInnerId0 < occaInnerDim0; ++occaInnerId0){
      }}}
}}}
\end{lstlisting}
\vspace{-2mm}

With the difference in work-item generation, having a work-item return in the middle of a kernel is problematic.
Although it is bad practice to have single work-items \tf{return} inside an OpenCL/CUDA kernel, we include \tf{occaInnerReturn} to allow GPU kernel work-items to \tf{return} while threads from OpenMP call a \tf{continue}.
Barriers needed to sync work-items will split loop groups, as seen in \refLst{lst:barrierSplitting}, which would require an OpenMP thread to \tf{continue} in the successive loops if intended to call \tf{return}.

\subsection{Threaded barriers}

Enabling the use of memory barriers is critical with memory latency becoming a more crucial issue due to new GPU architectures focusing on a greater stream processor to bandwidth ratio.
For optimized codes, the use of faster memory types, such as local and register memory, is indicated.
The thread-based granularity of GPU programming requires both, local and global barriers \opt{syncthreads} as an important part of the GPU programming model.
Many optimized kernels in GPUs make use of manual caching with local \opt{shared} memory requiring barrier-support in \occa \cite{lee2010debunking}.

For languages such as OpenMP that run work-groups in parallel but work-items in serial, intermediate barriers in the inner-loops have no synchronization like OpenCL or CUDA that run work-items in parallel.
Serial work-items require a split in inner-loops, as seen in \refLst{lst:barrierSplitting}, to go through each iteration statement prior of the barrier.
The \occa barriers are given by \tf{occaBarrier(occaLocalMemFence)} and \tf{occaBarrier(occaGlobalMemFence)} for local and global memory respectively.
Expansions to the keywords can be found in \refTab{tab:barriers} of the appendix.

\setCode{occa,none,\scriptsize}
\vspace{3mm}
\begin{lstlisting}[caption={
  Handling barriers in \occa requires the loop splitting seen between (2.1) and (2.2) in this code listing.
  OpenMP handles outer-loops in parallel and inner-loops serially, needing a break only in inner-loops.
},label={lst:barrierSplitting}]
// Loop grouping (1)
// Shared memory defined here

occaInnerFor2{ // Loop grouping (2.1)
  occaInnerFor1{
    occaInnerFor0{
}}}

occaBarrier(occaLocalMemFence);

occaInnerFor2{ // Loop grouping (2.2)
  occaInnerFor1{
    occaInnerFor0{
}}}
\end{lstlisting}
\vspace{-2mm}

\subsection{Registers and register arrays}

As seen in the \refLst{lst:barrierSplitting}, barriers cause a break in inner-loops.
Having explicit loops when using OpenMP requires memory to be managed through discontinuous scopes.
In other words, variable definitions in (2.1) are lost in scope and do not carry through to (2.2) in \refLst{lst:barrierSplitting}, making register initialization non-trivial.
Apart from scope visibility, overwriting becomes an issue with private variables defined outside of inner-loops.
In OpenMP, each outer-loop is handled in parallel while inner-loops are executed serially, causing private register values to be overwritten.
If the variable \tf{reg} in \refLst{lst:memoryTypes} would be initialized as a regular variable, the value of \tf{reg} would get overwritten by \tf{occaGlobalId0} in each iteration.

We introduce two \occa keywords for accessing registers and register arrays based on a work-item's ID: \tf{occaPrivate(type, name)} for and \tf{occaPrivateArray(type, name, size)}.
For GPU \occa, the macros expand as a regular register or register array initialization.
For OpenMP code, however, the macro expansion initializes an array with memory allocated for each work-item \opt{thread} in a buffer.
Operator overloading hides both, the work-item ID and required index in the array, treating \tf{occaPrivate} and \tf{occaPrivateArray} types as normal arrays.
Similarly, device-dependent implementations can be done with the use of the \occa keywords \tf{occaCPU}, \tf{occaGPU}, located in \refTab{tab:platformVariables} of the appendix.
Platform-dependent implementations can be managed with the use of \tf{occaOpenMP}, \tf{occaOpenCL}, and \tf{occaCUDA} which are also found in \refTab{tab:platformVariables} of the appendix.

\setCode{occa,none,\scriptsize}
\vspace{3mm}
\begin{lstlisting}[caption={
  The different memory types used in \occa kernels can be seen in this code listing.
  \tf{occaPrivate} and \tf{occaPrivateArray} variables remove write-conflicts that might appear in the serial inner-loop execution when using OpenMP-mode.
},label={lst:memoryTypes}]
occaKernel void kernelExample(occaKernelInfoArg, ...){
  occaOuterFor0{
    occaPrivate(int, reg);              // Register      : int reg;
    occaPrivateArray(int, regArray, 2); // Register Array: int regArray[2];

    // . . .

    occaInnerFor0{
      reg = occaGlobalId0; // reg would normally be overwritten by the loop
      regArray[0] = 0;     // regArray would also normally be overwritten
      regArray[1] = 1;
      // . . .
    }

    occaBarrier(occaLocalMemFence);

    occaInnerFor0{
      int i = reg;         // Allocating registers normally for only one scope
      int d = regArray[0];
      // . . .
    }
  }
}
\end{lstlisting}
\vspace{-2mm}


\section{Numerical Examples}
\label{sec:methods}

While \occa is made as a general-purpose portable language, our focus is to apply \occa for high performance implementations of numerical methods.
For this paper, we describe three different numerical methods implementing \occa and give performance results on different architectures.
We first present a finite difference (FD) example together with pseudocode, \occa kernel and \occa host code.
For performance results using \occa, we introduce two distinct numerical methods, one using a spectral element method (SEM) and the other a discontinuous Galerkin method (DG).

Kernel performance in this paper is compared between CPU architectures separately from GPU architectures with the GFLOP count (1e9 floating point operations) and bandwidth (memory transfer rate) as metrics.
Results between the distinct CPU and GPU architectures on competing platforms (OpenMP vs OpenCL-CPU) and (OpenCL-GPU vs CUDA) are shown separately for each project.
For the CPU architecture, OpenMP is compared with both, Intel's OpenCL platform and AMD's OpenCL platform on an Intel i7-3930K.
For the GPU architecture, NVIDIA's OpenCL platform and CUDA are compared on NVIDIA's Titan together with AMD's OpenCL platform on the AMD Radeon 7990 card.
Only one GPU processor is used on the dual-chip GPUs.

\subsection{Finite Difference}

The first \occa example is a finite difference code with a brief PDE and discretization summary.
Rather than optimization, the example's purpose is to show the construction of a simple \occa project with both, host and device code.
Performance studies on \occa-supported languages, however, are the focus of the two projects involving SEM and DG which are optimized for performance.
The SEM and DG codes are discussed further in the section.

\subsubsection{Governing Equations}

The 2D time-dependent PDE governing the isotropy wave equation is given by

\[u_{tt} = u_{xx} + u_{yy}\]

where $u$ is a scalar field representing acoustic pressure.
Sound speed in the material is of constant value 1 for simplicity.

\subsubsection{Discretization}

The 2D domain is a standard square $[-1,1]\times[-1,1]$ and discretized into a structured grid of \tf{w} node width and \tf{h} node height.
A first order second derivative stencil is used for stepping in time ($\{\frac{1}{\D t^2}, -\frac{2}{\D t^2}, \frac{1}{\D t^2}\}$).
The spatial differential discretization is given by a stencil of size $2r + 1$ in each dimension, where $r$ is the stencil radius.
The discrete stencil operator becomes

\[u_{n+1}(x_i,y_j) = (-2u_{n} + u_{n-1})
                     -\underset{\ds u_{xx}}{\underbrace{\sum_{k = -r}^r \omega_ku_{n}(x_{i+k},y_j)}}
                     -\underset{\ds u_{yy}}{\underbrace{\sum_{k = -r}^r \omega_ku_{n}(x_i,y_{j+k})}},\]

where $u_n$ is the solution stored at $t = t_n$ using a $2r$ order FD stencil, containing $2r + 1$ nodes in each dimension.
For simplicity in this example, no memory retrieval optimizations are made.
To find an optimized GPU finite difference code for the wave equation, we refer the reader to \cite{micikevicius20093d}.

\begin{algorithm}
\caption{Finite Difference Pseudocode}
\label{alg:fdPseudocode}
\begin{algorithmic}[1]

\ForAll {$(wg_i,wg_j)$}                                                 \hfill {\footnotesize\{ For each work-group in a 2D grid \}}
  \ForAll {$(i,j)$}                                                     \hfill {\footnotesize\{ For each thread in the 2D work-group\}}
    \State {$lap \leftarrow 0$}
    \For {$-r \leq k \leq r$}                                           \hfill {\footnotesize\{ For all nodes in the 1D stencil \}}
      \State {$lap \leftarrow lap + \omega_k u_n(i + k,j) + \omega_k u_n(i, j + k)$} \hfill {\footnotesize\{ Node contributions \}}
    \EndFor
    \State {$u_{n+1}(i,j) \leftarrow (-2u_n(i,j) + u_{n-1}(i,j) - \D t^2*lap)$}  \hfill {\footnotesize\{ Store solution at $t = t_{n+1}$\}}
  \EndFor
\EndFor
\end{algorithmic}
\end{algorithm}

The discrete Laplacian in the finite difference method seen in \refAlg{alg:fdPseudocode} iterates over a 1D stencil for each dimension.
The \occa kernel code for the pseudocode in \refAlg{alg:fdPseudocode} is shown in \refLst{lst:fdCode}.

\setCode{occa,none,\scriptsize}
\vspace{3mm}
\begin{lstlisting}[caption={
  The \occa finite difference kernel for the acoustic wave equation can be seen in this code listing.
  The pseudocode in \refAlg{alg:fdPseudocode} can be used for reference with this code listing.
},label={lst:fdCode}]
// External variables are defined as compiler directives
occaKernel void fd2d(occaKernelInfoArg,
                     occaPointer double *u1,
                     occaPointer double *u2,
                     occaPointer double *u3){
  occaOuterFor1{
    occaOuterFor0{
      occaInnerFor1{
        occaInnerFor0{
          const int i = occaGlobalId0;
          const int j = occaGlobalId1;

          const int id = j*w + i;      // w = nodes in the x direction
                                       // h = nodes in the y direction
          if( (i < w) && (j < h) ){    // Bounds check
            double lap = 0.0;

            const double r_u1 = u1[id]; // Global to register memory
            const double r_u2 = u2[id];

            for(int k = -r; k <= r; k++){
              const int nX = (i + k + w) % w; // Periodic Boundary
              const int nY = (j + k + h) % h; // Periodic Boundary

              lap += weight[r + k]*u1[j*w + nX]  // Iterate horizontal nodes
                  +  weight[r + k]*u1[nY*w + i]; // Iterate vertical nodes
            }

            u3[id] = (-2*r_u1 + r_u2 - dt*dt*lap);
          }
      }}
  }}
}
\end{lstlisting}
\vspace{-2mm}

The work-group/work-item programming style in \occa requires a bounds check on the thread IDs as seen in \refLst{lst:fdCode}.
In addition to the kernel code seen in \refLst{lst:fdCode}, we provide code snippets in \refLst{lst:fdHostCode} to show the \occa host API used in the FD example.

\setCode{occa,none,\scriptsize}
\vspace{3mm}
\begin{lstlisting}[caption={
  The \occa host code used for the finite difference code is in this listing
},label={lst:fdHostCode}]
occa helper;
occaMemory o_u1, o_u2, o_u3;
occaKernel fd2d;

// . . .

void setupSolver(){
  // Setup occa helper
  // Model can be: "OpenCL", "CUDA", "OpenMP"
  helper.setup(model.c_str(), plat, dev);

  // Allocate device memory
  o_u1 = helper.createBuffer(u1); // u1, u2, u3 are 0 matrices of size w by h
  o_u2 = helper.createBuffer(u1);
  o_u3 = helper.createBuffer(u2);

  // Add kernel definitions
  string defs;

  helper.addDefine(defs, "r" , stencilRadius);
  helper.addDefine(defs, "w" , width);
  helper.addDefine(defs, "h" , height);
  helper.addDefine(defs, "dx", dx);
  helper.addDefine(defs, "dt", dt);

  // Build kernel
  size_t dims      = 2;
  size_t local[2]  = {16, 16};
  size_t global[2] = {local[0]*( (width  + local[0] - 1)/local[0] ),
                      local[1]*( (height + local[1] - 1)/local[1] )};

  fd2d = helper.buildKernel("fd2d.occa", "fd2d", defs);
  fd2d.setThreadArray(global, local, dims);
}

void timestep(){
  currentTime += dt;

  fd2d(o_u1, o_u2, o_u3, currentTime);

  // Shift timestep solutions to: u1 <- u2 <- u3
  o_u1.swap(o_u2); // Swaps memory handle between u1 and u2
  o_u2.swap(o_u3); // Swaps memory handle between u2 and u3
}

void main(){
  loadVariables(); // Load all variables (model, width, height, ...)
  setupSolver();

  while(true)
    timestep();
}
\end{lstlisting}
\vspace{-2mm}

\subsubsection{Finite Difference Performance}\fnl

\def\plotWidth{200pt}
\def\plotHeight{220pt}

\setPlotWidth{\plotWidth}
\setPlotHeight{\plotHeight}
\setPlotXTitle{Stencil Size}
\setPlotYTitle{MNodes/s}

\setPlotLegendPos{north east}

\begin{figure}[h!]
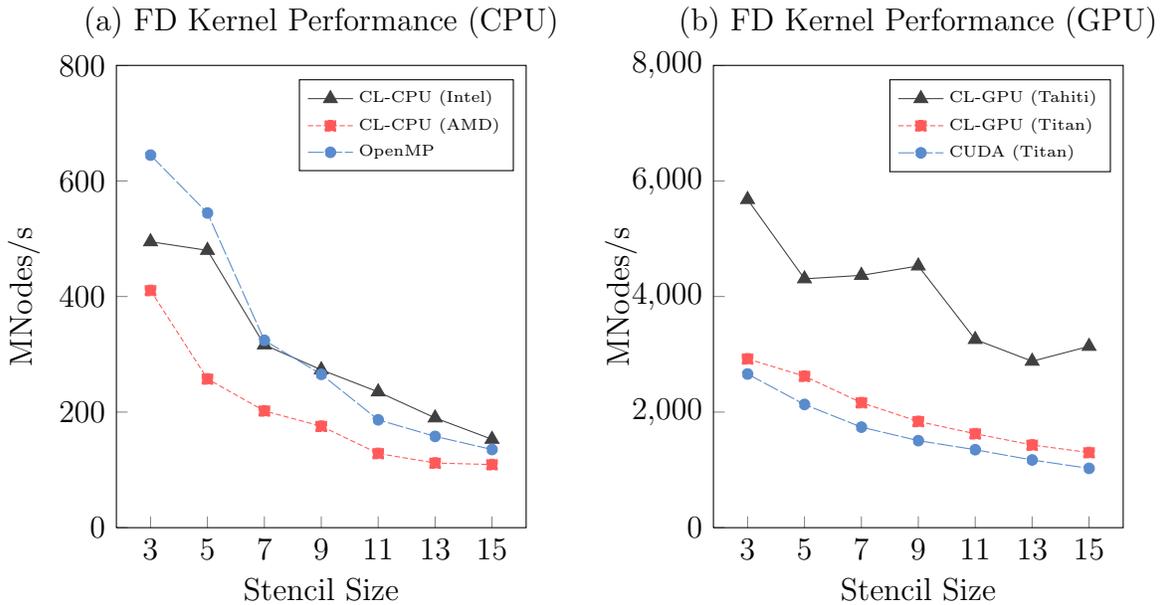

\begin{tabular}{cc}
\setPlotTitle{(a) FD Kernel Performance (CPU)}

\clearPlotSettings
\appendPlotSettings{ymin=0}
\appendPlotSettings{ymax=800}

\setPlotFile{fdMNodes.csv}

\setPlotXLabels{3,5,7,9,11,13,15}
\setPlotLegendEntries{CL-CPU (Intel), CL-CPU (AMD), OpenMP}
\setPlotValueColumns{CL-Intel-CPU, CL-AMD-CPU, OpenMP}
\setPlotValueColumnCount{3}

\hspace{-23mm}
\plotYLine
&
\hspace{-27mm}

\setPlotTitle{(b) FD Kernel Performance (GPU)}

\clearPlotSettings
\appendPlotSettings{ymin=0}
\appendPlotSettings{ymax=8000}

\setPlotFile{fdMNodes.csv}

\setPlotXLabels{3,5,7,9,11,13,15}
\setPlotLegendEntries{CL-GPU (Tahiti), CL-GPU (Titan), CUDA (Titan)}
\setPlotValueColumns{CL-AMD-GPU, CL-NVIDIA-GPU, CUDA}

\plotYLine
\end{tabular}
\caption{
  Performance in MNodes/s of the FD kernel in \refLst{lst:fdCode} are shown in (a) and (b).
  Results shown in figure (a) were obtained on three CPU platforms on an Intel i7-3930K 6-core processor.
  Results shown in figure (b) were obtained on an NVIDIA Titan card and AMD Radeon 7970.
}
\label{fig:fdTable}
\end{figure}

The performance of finite difference implementations is usually given by the millions of nodes processed per second.
Taking the standard performance metric in finite difference implementations, \refFig{fig:fdTable} uses MNodes/s to gauge performance.
Figure (a) shows similar performance between OpenMP and Intel's OpenCL platform, while AMD's OpenCL platform does not seem to vectorize well.
Similar results are seen in following CPU performance results in \refFig{fig:cpuSemTable} and \refFig{fig:cpuDGTable}.
As for GPU comparisons, figure (b) results show OpenCL performs better than CUDA in NVIDIA's Titan card while OpenCL on AMD's Tahiti outperforms the Titan results for this case.

\subsection{Spectral Element Methods}

In our second numerical method example, we focus on the screened Coulomb potential elliptic PDE for the spectral element example,

\sEq\label{eq:semPDE}
-\div(\kappa \grad u) + \alpha u = f,\ \text{in } \O\subset\R^3,
\eEq

where we solve for temperature $u(x)$ given a source $f(x)$ and material-dependent coefficients $\kappa(x)$ and $\alpha(x)$.

\subsubsection{Discretization}

We consider a spectral element method formulation using hexahedral elements $D^e$ on the discretized domain $\O_h = \bigcup_{e=0}^E D^e$.
The weak form of \refEq{eq:semPDE} is then given by,

\sEq \label{eq:semOperator}
  \underset{\text{Stiffness Operator}}{\underbrace{\wvs\ip{\kappa\grad u}{\grad v}_\O}}
+ \underset{\text{Mass Operator}}{\underbrace{\wvs\ip{\alpha u}{v}_\O}} = \ip{f}{v}_\O,\ u,v\in V
\eEq

where $V$ is the space of continuous piecewise polynomials of degree $N$ on $\O_h$.
Making use of the tensor-product structure of hexahedral elements, the basis functions on each element $\phi_{abc}$ become a tensor product of Lagrange polynomials on the 1D Gauss-Lobatto-Legendre nodes $\hat\phi_{i}$ ($i = 0, 1, \ldots, N$).
The tensor expansion results in the basis

\[\phi_{abc}(x,y,z) = \hat\phi_{a}(x)\hat\phi_{b}(y)\hat\phi_{c}(z),\]

where $x(r,s,t),\ y(r,s,t),\ z(r,s,t)$ are mapped from the reference hexahedral element defined by $\hat D = [-1,1]\times[-1,1]\times[-1,1]$ under the coordinates $r, s, t$.
Using Lagrange basis on quadrature nodes gives a discrete orthogonality on an element $D^e$, leading to

\sAr
\ip{\phi_{abc}}{\phi_{ijk}}_{D^e} & \approx \ds J^e\left(\sum_{m,n,o = 0}^{N} w_{mno}\phi_{abc}(r_m,s_n,t_o)\phi_{ijk}(r_m,s_n,t_o)\right) \nl
                                  & =       \ds J^e\left(\sum_{m,n,o = 0}^{N} w_{mno}\delta_{aim}\delta_{bjn}\delta_{cko}          \right) \nll
                                  & =       \left\{\sArr{ll} J^ew_{abc} & abc = ijk\\ 0 &\text{otherwise}\eArr\right.,
\eAr

where $J^e$ is the Jacobian from the reference mapping for element $e$, lumping the discrete mass operator as a diagonal matrix of quadrature weights.
Similarly, quadrature weights, $w_{abc}$, are formed from a tensor of 1D quadrature weights, $\hat{w}_a\hat{w}_b\hat{w}_c$.
A second simplification we make also comes from the tensor product formation of the basis functions, giving us

\[
\vp{}{r}\phi_{abc}\at_{(r_m,s_n,t_o)}
  = \left\{
      \sArr{ll}
       \phi_a'(r_m) & bc = no\nl
       0            & \text{otherwise}
      \eArr
    \right.,
\]

Using chain-rule,

\[u_x(r,s,t) = r_xu_r(r,s,t) + s_xu_s(r,s,t) + t_xu_t(r,s,t),\]

we simplify the discrete stiffness operator \refP{eq:semOperator} to

\[\sArr{ll}
  \ds\int_{\hat D}
  &\ip{\kappa\grad u}{\grad \phi_{ijk}}_{D^e}
    \nl
    &\approx \ds\int_{\hat D} J^e
       \left(\vT{\kappa u_r\wvs, \kappa u_s\wvs, \kappa u_t\wvs}
         \underset{G^T}{\underbrace{
         \left[\sArr{ccc}
         r_x^e & r_y^e & r_z^e \nl
         s_x^e & s_y^e & s_z^e \nl
         t_x^e & t_y^e & t_z^e \nl
         \eArr\right]}}
       \underset{G}{\underbrace{
         \left[\sArr{ccc}
         r_x^e & s_x^e & t_x^e \nl
         r_y^e & s_y^e & t_y^e \nl
         r_z^e & s_z^e & t_z^e \nl
         \eArr\right]}}
         \v{\vp{}{r}\phi_{ijk}\wvs, \vp{}{s}\phi_{ijk}\wvs, \vp{}{t}\phi_{ijk}\wvs}\right) \nl
    &= J^e\left(\ds\sum_{m=0}^N \vT{
              \hat u^e_{mjk}\hat{\phi}'_i(r_m),
              \hat u^e_{imk}\hat{\phi}'_j(s_m),
              \hat u^e_{ijm}\hat{\phi}'_k(t_m)}\right)
           \hat G
          \left(\ds\sum_{m=0}^N \v{
              \hat{\phi}'_i(r_m),
              \hat{\phi}'_j(s_m),
              \hat{\phi}'_k(t_m)}\right)
\eArr\]

where $r^e_*, s^e_*, t^e_*$ are the reference mapping geometric factors for element $e$, $\hat G = G^TG$ and
$\hat u^e_{abc} = \kappa^e(r_a, s_b, t_c) u^e(r_a, s_b, t_c)$.
The geometric factors in $\hat G$ are precomputed for each elemental node.

The parallel implementation is derived from \cite{fischer1991parallel} using distinct global-local node numbering.
We refer to \cite{deville2002high} for further explanation on high-order spectral element methods.

\subsubsection{Spectral Element Performance}\fnl

The screened Coulomb potential PDE was numerically solved by a spectral element method discretization and preconditioned conjugate gradient linear solver (PCG).
Besides the preconditioner choice, the most computational-intensive routine in the PCG iterations comes from the matrix application, or the SEM operator.
We record performance of one \occa kernel which applies SEM operator seen in \refEq{eq:semOperator} for the different CPU and GPU platforms.

\setPlotWidth{\plotWidth}
\setPlotHeight{\plotHeight}
\setPlotXTitle{Polynomial Order}
\setPlotYTitle{GFLOPS}

\setPlotLegendPos{north west}

\begin{figure}[h!]
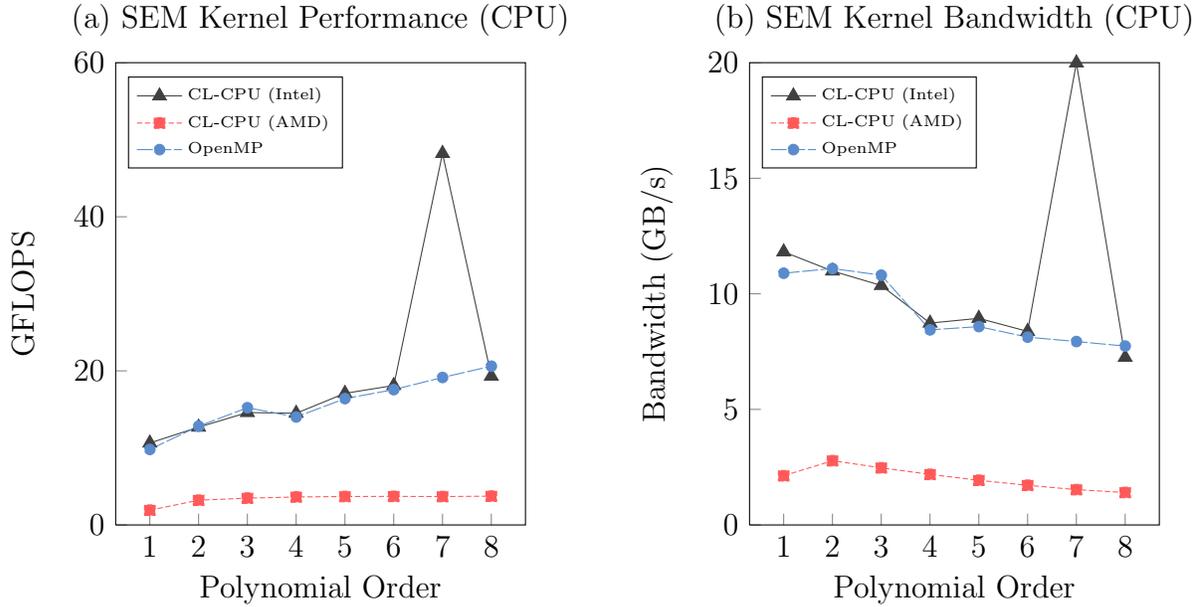

\begin{tabular}{cc}
\setPlotTitle{(a) SEM Kernel Performance (CPU)}

\clearPlotSettings
\appendPlotSettings{ymin=0}
\appendPlotSettings{ymax=60}
\setPlotYTitle{GFLOPS}

\setPlotFile{semGFLOPS.csv}

\setPlotXLabels{1,2,3,4,5,6,7,8}
\setPlotLegendEntries{CL-CPU (Intel), CL-CPU (AMD), OpenMP}
\setPlotValueColumns{CL-Intel-CPU, CL-AMD-CPU, OpenMP}
\setPlotValueColumnCount{3}

\hspace{-25mm}
\plotYLine
&
\hspace{-25mm}

\setPlotTitle{(b) SEM Kernel Bandwidth (CPU)}

\clearPlotSettings
\appendPlotSettings{ymin=0}
\appendPlotSettings{ymax=20}
\setPlotYTitle{Bandwidth (GB/s)}

\setPlotFile{semBW.csv}

\setPlotXLabels{1,2,3,4,5,6,7,8}
\setPlotLegendEntries{CL-CPU (Intel), CL-CPU (AMD), OpenMP}
\setPlotValueColumns{CL-Intel-CPU, CL-AMD-CPU, OpenMP}

\plotYLine
\end{tabular}
\caption{
  Performance results in figures (a) and (b) were taken from the \occa SEM discrete operator in \cite{moon2014brainnek}.
  Results shown in both figures were obtained on an Intel i7-3930K 6-core processor using OpenMP and both, Intel and AMD OpenCL platforms.
  Figures (a) and (b) graph GFLOP performance and bandwidth usage respectively.
}
\label{fig:cpuSemTable}
\end{figure}

\setPlotWidth{\plotWidth}
\setPlotHeight{\plotHeight}
\setPlotXTitle{Polynomial Order}

\setPlotLegendPos{north west}

\begin{figure}[h!]
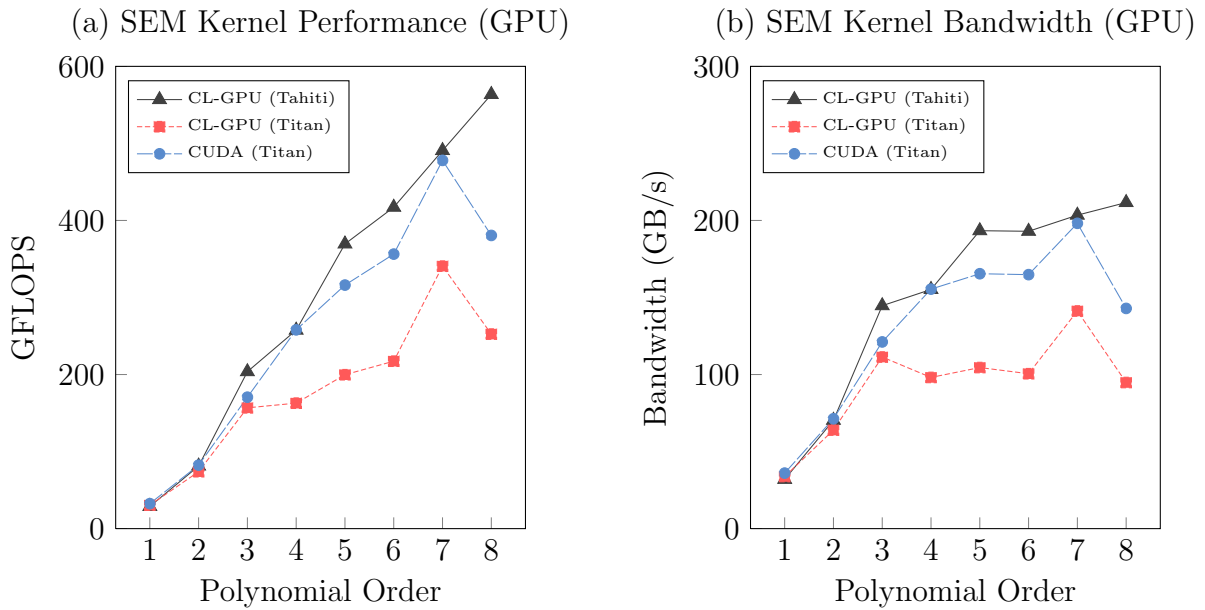

\begin{tabular}{cc}
\setPlotTitle{(a) SEM Kernel Performance (GPU)}

\clearPlotSettings
\appendPlotSettings{ymin=0}
\appendPlotSettings{ymax=600}
\setPlotYTitle{GFLOPS}

\setPlotFile{semGFLOPS.csv}

\setPlotXLabels{1,2,3,4,5,6,7,8}
\setPlotLegendEntries{CL-GPU (Tahiti), CL-GPU (Titan), CUDA (Titan)}
\setPlotValueColumns{CL-AMD-GPU, CL-NVIDIA-GPU, CUDA}
\setPlotValueColumnCount{3}

\hspace{-25mm}
\plotYLine
&
\hspace{-25mm}

\setPlotTitle{(b) SEM Kernel Bandwidth (GPU)}

\clearPlotSettings
\appendPlotSettings{ymin=0}
\appendPlotSettings{ymax=300}
\setPlotYTitle{Bandwidth (GB/s)}

\setPlotFile{semBW.csv}

\setPlotXLabels{1,2,3,4,5,6,7,8}
\setPlotLegendEntries{CL-GPU (Tahiti), CL-GPU (Titan), CUDA (Titan)}
\setPlotValueColumns{CL-AMD-GPU, CL-NVIDIA-GPU, CUDA}

\plotYLine
\end{tabular}
\caption{
  Performance results in figures (a) and (b) were taken from the \occa SEM discrete operator in \cite{moon2014brainnek}.
  Results were obtained using CUDA and OpenCL on an NVIDIA Titan card and OpenCL on an AMD Radeon 7970.
  Figures (a) and (b) graph GFLOP performance and bandwidth usage respectively.
}
\label{fig:gpuSemTable}
\end{figure}

Results seen in \refFig{fig:cpuSemTable} show a large gap between Intel's and AMD's OpenCL platform on an Intel processor (i7-3930K).
Because of the 8-way vectorization Intel is using, there is a large gain from using explicit for-loops in the discrete operator with $N = 7$.
Aside from the outlier, Intel's OpenCL performs similar to OpenMP.

Shown in \refFig{fig:gpuSemTable} are comparisons between OpenCL (ver. 1.2) and CUDA (arch 3.5) using a GTX Titan card showing a difference in performance.
Alongside are results using OpenCL on one of the chips of an AMD 7990.
Bandwidth peak in \refFig{fig:gpuSemTable} and later seen in \refFig{fig:gpuDGTable} can be explained by the use of shared memory and caching due to well aligned memory.
For further reference in the application and methodology, refer to \cite{moon2014brainnek}.

\subsection{Discontinuous Galerkin for Shallow Water Equations}

In our third and last numerical example, we consider a high-order discontinuous Galerkin (DG) numerical solver for the shallow water equations (SWE).
DG methods are used in a broad scope of applications, such as Maxwell's equations \cite{klockner2009nodal}, wave propagation \cite{burstedde2010extreme} and computational fluid dynamics solvers \cite{gerber2013benchmarking}.
The SWE example uses a multi-step method Adam-Bashforth scheme with local time-stepping for time integration.
The trial and test space are spanned by nodal basis functions and are further explained in \cite{hesthaven2008nodal}.

\subsubsection{Shallow Water Governing Equations}

The 2D shallow water equations in conservative form are

\sEq \label{eq:dgPDE}
\vp{Q}{t} + \vp{F}{x} + \vp{G}{y} = S,
\eEq

where $Q$ is the vector of conservative variables, $F$ and $G$ are nonlinear flux vectors, and S is a source term.
Variables from \refEq{eq:dgPDE} are defined as

\[Q = \vb{h, hu, hv},\ F = \vb{hu, hu^2 + \frac{gh^2}{2}, huv},\ G = \vb{hv, huv, hv^2 + \frac{gh^2}{2}},\ S = \vb{0, -gh\ts\vp{B}{x}\\[-4mm], -gh\ts\vp{B}{y}},\]

where $h$ is the water depth, $u,v$ are the velocity components.
Bathymetry is represented by $B$, describing the distribution of the ocean bed topography, and acceleration from gravity is given by $g$.

\subsubsection{Discontinuous Galerkin Discretization}

The domain $\O\subset\R^2$ is discretized and partitioned into non-overlapping and conforming triangular elements $D^e$ such that $\O \approx \O_h = \bigcup_{e=0}^E D^e$.
We find an approximated solution $Q_H\in P^N(D^e)$, the space of discontinuous piecewise polynomials of degree $N$, using the weak formulation on each element

\[\ip{\vp{Q_H}{t}}{v}_{D^e} +\ \ip{\vp{F}{x}}{v}_{D^e} +\ \ip{\vp{G}{y}}{v}_{D^e} = \ip{S}{v}_{D^e},\ v\in P^N(D^e). \]

The fluxes obtained by integration by parts, results in the weak form

\[\ip{\vp{Q_H}{t}}{v}_{D^e} =\
                      \underset{\text{Volume Kernel}}{\underbrace{\ip{F}{\vp{v}{x}}_{D^e} +\ \ip{G}{\vp{v}{y}}_{D^e} + \ip{S}{v}_{D^e}}} +
                     -\underset{\text{Surface Kernel}}{\underbrace{\ip{F^*n_x + G^*n_y}{v}_{\partial D^e}\vphantom{\ds\ip{\vp{G}{y}}{v}_{D^e}}}},\]

for $v\in P^N(D^e)$.
For further explanation of the SWE application, we refer to \cite{gandham2014swe}.
We refer to the GPU Computing Gems Jade Edition \cite{klockner2012jade} for an OpenCL example and for further DG explanation we refer to \cite{hesthaven2008nodal}.

\setPlotWidth{\plotWidth}
\setPlotHeight{\plotHeight}
\setPlotXTitle{Polynomial Order}

\begin{figure}[h!]
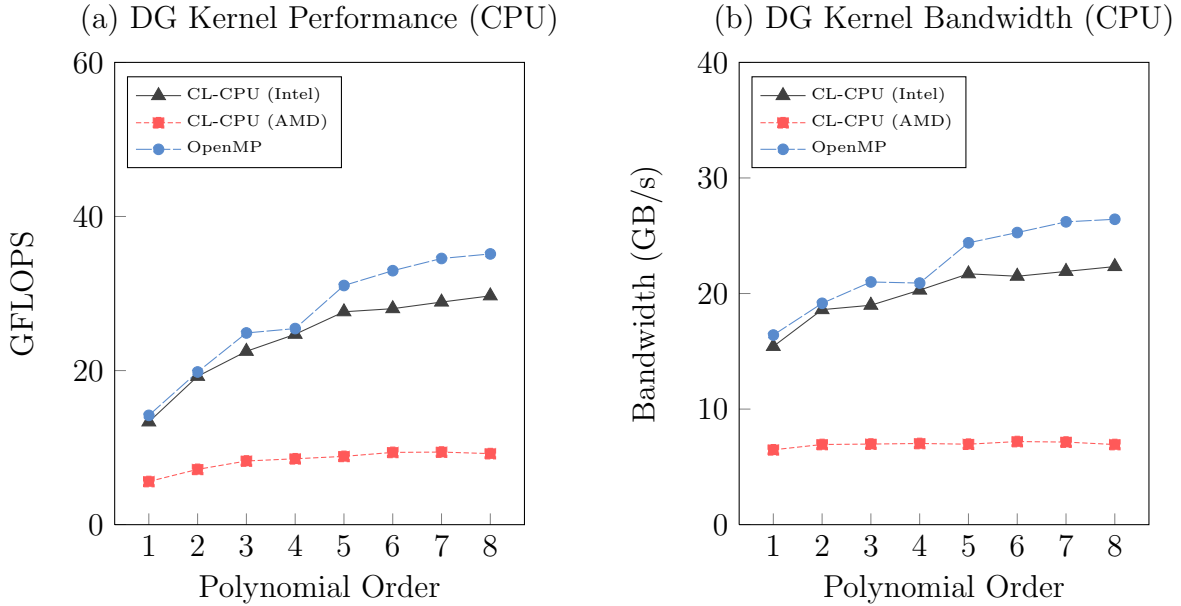

\begin{tabular}{cc}
\setPlotTitle{(a) DG Kernel Performance (CPU)}

\clearPlotSettings
\appendPlotSettings{ymin=0}
\appendPlotSettings{ymax=60}
\setPlotYTitle{GFLOPS}

\setPlotFile{dgVolGFLOPS.csv}

\setPlotXLabels{1,2,3,4,5,6,7,8}
\setPlotLegendEntries{CL-CPU (Intel), CL-CPU (AMD), OpenMP}
\setPlotValueColumns{CL-Intel-CPU, CL-AMD-CPU, OpenMP}
\setPlotValueColumnCount{3}

\hspace{-25mm}
\plotYLine
&
\hspace{-25mm}

\setPlotTitle{(b) DG Kernel Bandwidth (CPU)}

\clearPlotSettings
\appendPlotSettings{ymin=0}
\appendPlotSettings{ymax=40}
\setPlotYTitle{Bandwidth (GB/s)}

\setPlotFile{dgVolBW.csv}

\setPlotXLabels{1,2,3,4,5,6,7,8}
\setPlotLegendEntries{CL-CPU (Intel), CL-CPU (AMD), OpenMP}
\setPlotValueColumns{CL-Intel-CPU, CL-AMD-CPU, OpenMP}

\plotYLine
\end{tabular}
\caption{
  Performance results in figures (a) and (b) were taken from the DG volume integration kernel appearing in \cite{gandham2014swe}.
  Results shown in both figures were obtained on an Intel i7-3930K 6-core processor using OpenMP and both, Intel and AMD OpenCL platforms.
  Figures (a) and (b) graph GFLOP performance and bandwidth usage respectively.
}
\label{fig:cpuDGTable}
\end{figure}

\setPlotWidth{\plotWidth}
\setPlotHeight{\plotHeight}
\setPlotXTitle{Polynomial Order}

\begin{figure}[h!]
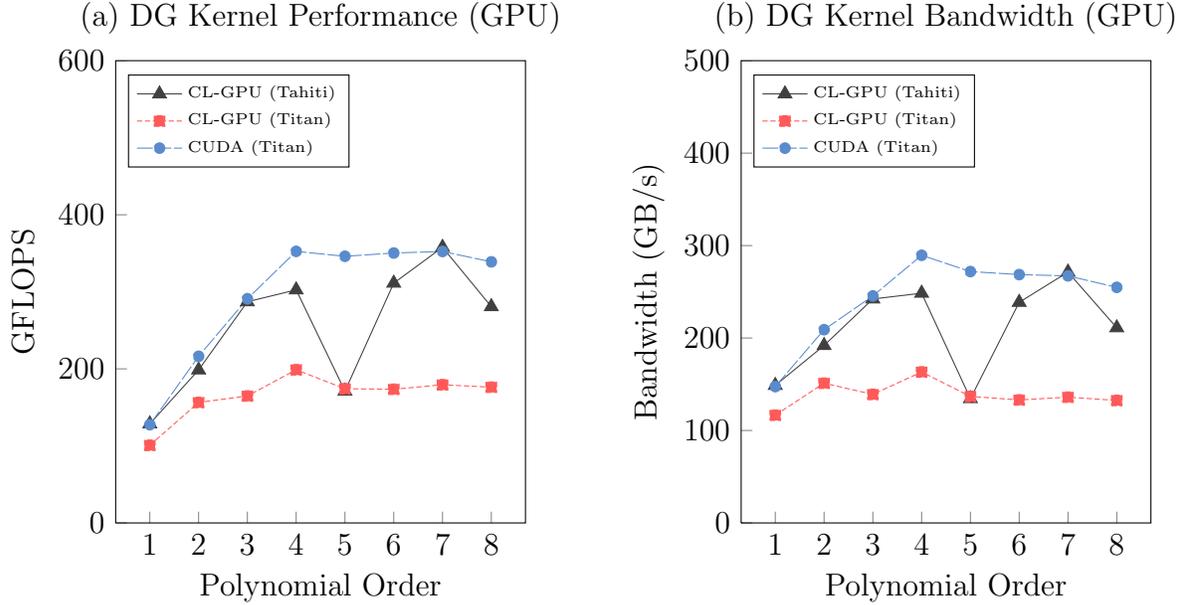

\begin{tabular}{cc}
\setPlotTitle{(a) DG Kernel Performance (GPU)}

\clearPlotSettings
\appendPlotSettings{ymin=0}
\appendPlotSettings{ymax=600}
\setPlotYTitle{GFLOPS}

\setPlotFile{dgVolGFLOPS.csv}

\setPlotXLabels{1,2,3,4,5,6,7,8}
\setPlotLegendEntries{CL-GPU (Tahiti), CL-GPU (Titan), CUDA (Titan)}
\setPlotValueColumns{CL-AMD-GPU, CL-NVIDIA-GPU, CUDA}
\setPlotValueColumnCount{3}

\hspace{-25mm}
\plotYLine
&
\hspace{-25mm}

\setPlotTitle{(b) DG Kernel Bandwidth (GPU)}

\clearPlotSettings
\appendPlotSettings{ymin=0}
\appendPlotSettings{ymax=500}
\setPlotYTitle{Bandwidth (GB/s)}

\setPlotFile{dgVolBW.csv}

\setPlotXLabels{1,2,3,4,5,6,7,8}
\setPlotLegendEntries{CL-GPU (Tahiti), CL-GPU (Titan), CUDA (Titan)}
\setPlotValueColumns{CL-AMD-GPU, CL-NVIDIA-GPU, CUDA}

\plotYLine
\end{tabular}
\caption{
  Performance results in figures (a) and (b) were taken from the DG volume integration kernel appearing in \cite{gandham2014swe}.
  Results shown were obtained using CUDA and OpenCL on an NVIDIA Titan card and OpenCL on an AMD Radeon 7970.
  Figures (a) and (b) graph GFLOP performance and bandwidth usage respectively.
}
\label{fig:gpuDGTable}
\end{figure}

\subsubsection{Discontinuous Galerkin Performance}\fnl

We chose to profile the volume kernel for this performance display due to being the most computational intensive kernel in terms of floating-point operations.
Results shown in \refFig{fig:gpuDGTable} are obtained using optimal kernel tuning parameters.
The mesh used to compute the results seen in \refFig{fig:cpuDGTable} and \refFig{fig:gpuDGTable} contains 212,800 elements \cite{gandham2014swe}..

\refFIG{fig:cpuDGTable} shows a similar trend seen in the SEM results, showing OpenMP and Intel's OpenCL platform displays similar performance.
Intel's OpenCL platform and OpenMP operated similarly at low orders but outperformed in GFLOPS at high orders while AMD's OpenCL platform showed a fraction of the performance.
\refFIG{fig:gpuDGTable} results show CUDA outperformed OpenCL in the Titan GPU but OpenCL showed similar results on some of the polynomial orders.

\section{Concluding Remarks}

We introduced a method which combined elements from multi-threading and many-core parallel processing to create a unified multi-threading language.
Using this macro-based approach allows codes to differ in architecture and platform execution while providing the flexibility of supporting additional languages.
One novelty of \occa, aside from the unification of device languages, is the run-time compilation option, simplifying hybrid computing in parallel environments.
Performance profiles of the finite difference, spectral element and discontinuous Galerkin methods gave real-file applications utilizing \occa with platform portability without sacrificing performance.

\section{Acknowledgments}

The authors are grateful and wish to acknowledge the grants from Shell (Shell Agreement No. PT22584), DOE and ANL (ANL Subcontract No. 1F-32301 on DOE grant NO. DE-AC02-06CH11357), support from both, the Shell fellowship and AMD.

\section{Appendix: \OCCA Kernel Keywords}\label{sec:appendix}

Provided are \occa keywords categorized in tables by their use or purpose.
Each table provides similar \occa keywords on the left-most column.
Adjacent to the \occa keyword column is the OpenMP macro expansion, followed by the OpenCL macro expansion and the CUDA macro expansion on the right-most column.
\vspace{9mm}

{\scriptsize
\sMini\begin{center}\begin{tabular}{|l|l|l|l|}
  \hline
  \wvs OCCA                     & OpenMP                                                          & OpenCL                               & CUDA                                     \\ \hline
  \rcA \tf{occaInnerId0       } & \tf{                            } \wvs & \tf{get\_local\_id(0) } & \tf{threadIdx.x            } \\[1mm]
  \rcB \tf{occaInnerId1       } & \tf{                            } \wvs & \tf{get\_local\_id(1) } & \tf{threadIdx.y            } \\[1mm]
  \rcA \tf{occaInnerId2       } & \tf{                            } \wvs & \tf{get\_local\_id(2) } & \tf{threadIdx.z            } \\[1mm]
  \rcB \tf{occaOuterId0       } & \tf{                            } \wvs & \tf{get\_group\_id(0) } & \tf{blockIdx.x             } \\[1mm]
  \rcA \tf{occaOuterId1       } & \tf{                            } \wvs & \tf{get\_group\_id(1) } & \tf{blockIdx.y             } \\[1mm]
  \rcB \tf{occaGlobalId0      } & \tf{occaInnerId0                } \wvs & \tf{get\_global\_id(0)} & \tf{threadIdx.x            } \\[-1mm]
  \rcB \tf{                   } & \tf{+ occaInnerDim0*occaOuterId0} \wvs & \tf{                  } & \tf{+ blockIdx.x*blockDim.x} \\[1mm]
  \rcA \tf{occaGlobalId1      } & \tf{occaInnerId1                } \wvs & \tf{get\_global\_id(1)} & \tf{threadIdx.y            } \\[-1mm]
  \rcA \tf{                   } & \tf{+ occaInnerDim1*occaOuterId1} \wvs & \tf{                  } & \tf{+ blockIdx.y*blockDim.y} \\[1mm]
  \rcB \tf{occaGlobalId2      } & \tf{occaInnerId2                } \wvs & \tf{get\_global\_id(2)} & \tf{threadIdx.z            } \\[1mm]
  \hline
\end{tabular}
\captionof{table}{\occa keywords used to obtain a scope's work-group \opt{block} and work-item \opt{thread} ID.}
\label{tab:threadingIDs}
\end{center}\eMini\vspace{9mm}
}


{
\sMini
\begin{center}
\scriptsize
\begin{tabular}{|l|l|l|l|}
  \hline
  \wvs OCCA                     & OpenMP                                                          & OpenCL                                 & CUDA                                   \\ \hline
  \rcA \tf{occaInnerDim0      } & \tf{occaDims[0]                } \wvs & \tf{get\_local\_size(0) } & \tf{blockDim.x                 } \\[1mm]
  \rcB \tf{occaInnerDim1      } & \tf{occaDims[1]                } \wvs & \tf{get\_local\_size(1) } & \tf{blockDim.y                 } \\[1mm]
  \rcA \tf{occaInnerDim2      } & \tf{occaDims[2]                } \wvs & \tf{get\_local\_size(2) } & \tf{blockDim.z                 } \\[1mm]
  \rcB \tf{occaOuterDim0      } & \tf{occaDims[3]                } \wvs & \tf{get\_num\_groups(0) } & \tf{gridDim.x                  } \\[1mm]
  \rcA \tf{occaOuterDim1      } & \tf{occaDims[4]                } \wvs & \tf{get\_num\_groups(1) } & \tf{gridDim.y                  } \\[1mm]
  \rcB \tf{occaGlobalDim0     } & \tf{occaInnerDim0*occaOuterDim0} \wvs & \tf{get\_global\_size(0)} & \tf{occaInnerDim0*occaOuterDim0} \\[1mm]
  \rcA \tf{occaGlobalDim1     } & \tf{occaInnerDim1*occaOuterDim1} \wvs & \tf{get\_global\_size(1)} & \tf{occaInnerDim1*occaOuterDim1} \\[1mm]
  \rcB \tf{occaGlobalDim2     } & \tf{occaInnerDim2*occaOuterDim2} \wvs & \tf{get\_global\_size(2)} & \tf{occaInnerDim2              } \\[1mm]
  \hline
\end{tabular}
\captionof{table}{\occa keywords related to storing work-group \opt{block} and work-item \opt{thread} sizes.}
\label{tab:workSizes}
\end{center}\eMini\vspace{9mm}
}


{\scriptsize
\sMini\begin{center}\begin{tabular}{|l|l|l|l|}
  \hline
  \wvs OCCA                     & OpenMP                                                          & OpenCL                               & CUDA                                     \\ \hline
  \rcA \tf{occaInnerFor       } & \tf{occaInnerFor2 occaInnerFor1 occaInnerFor0} \wvs & \tf{} & \tf{} \\[1mm]
  \rcB \tf{occaInnerFor0      } & \tf{for(occaInnerId0 = 0;                    } \wvs & \tf{} & \tf{} \\[-1mm]
  \rcB                          & \hspace{5mm}\tf{occaInnerId0 < occaInnerDim0;}      & \tf{} & \tf{} \\
  \rcB                          & \hspace{5mm}\tf{++occaInnerId0)              }      & \tf{} & \tf{} \\[1mm]
  \rcA \tf{occaInnerFor1      } & \tf{for(occaInnerId1 = 0;                    } \wvs & \tf{} & \tf{} \\[-1mm]
  \rcA                          & \hspace{5mm}\tf{occaInnerId1 < occaInnerDim1;}      & \tf{} & \tf{} \\
  \rcA                          & \hspace{5mm}\tf{++occaInnerId1)              }      & \tf{} & \tf{} \\[1mm]
  \rcB \tf{occaInnerFor2      } & \tf{for(occaInnerId2 = 0;                    } \wvs & \tf{} & \tf{} \\[-1mm]
  \rcB                          & \hspace{5mm}\tf{occaInnerId2 < occaInnerDim2;}      & \tf{} & \tf{} \\
  \rcB                          & \hspace{5mm}\tf{++occaInnerId2)              }      & \tf{} & \tf{} \\[1mm]
  \rcA \tf{occaOuterFor0      } & \tf{for(occaOuterId0 = 0;                    } \wvs & \tf{} & \tf{} \\[-1mm]
  \rcA                          & \hspace{5mm}\tf{occaOuterId0 < occaOuterDim0;}      & \tf{} & \tf{} \\
  \rcA                          & \hspace{5mm}\tf{++occaOuterId0)              }      & \tf{} & \tf{} \\[1mm]
  \rcB \tf{occaOuterFor1      } & \tf{for(occaOuterId1 = 0;                    } \wvs & \tf{} & \tf{} \\[-1mm]
  \rcB                          & \hspace{5mm}\tf{occaOuterId1 < occaOuterDim1;}      & \tf{} & \tf{} \\
  \rcB                          & \hspace{5mm}\tf{++occaOuterId1)              }      & \tf{} & \tf{} \\[1mm]
  \rcA \tf{occaOuterFor2      } & \tf{                                         } \wvs & \tf{} & \tf{} \\[1mm]
  \rcB \tf{occaGlobalFor0     } & \tf{occaOuterFor0 occaInnerFor0              } \wvs & \tf{} & \tf{} \\[1mm]
  \rcA \tf{occaGlobalFor1     } & \tf{occaOuterFor1 occaInnerFor1              } \wvs & \tf{} & \tf{} \\[1mm]
  \rcB \tf{occaGlobalFor2     } & \tf{occaInnerFor2                            } \wvs & \tf{} & \tf{} \\[1mm]
  \hline
\end{tabular}
\captionof{table}{\occa keywords related to explicitly displaying work-group \opt{block} and work-item \opt{thread} loop scopes.}
\label{tab:threadingDimensions}
\end{center}\eMini\vspace{9mm}
}


{\scriptsize
\sMini\begin{center}\begin{tabular}{|l|l|l|l|}
  \hline
  \wvs OCCA                     & OpenMP                                                          & OpenCL                               & CUDA                                     \\ \hline
  \rcA \tf{occaShared         } & \tf{                                                          } \wvs & \tf{\_\_local   } & \tf{\_\_shared\_\_  } \\[1mm]
  \rcB \tf{occaPointer        } & \tf{                                                          } \wvs & \tf{\_\_global  } & \tf{                } \\[1mm]
  \rcA \tf{occaConstant       } & \tf{                                                          } \wvs & \tf{\_\_constant} & \tf{\_\_constant\_\_} \\[1mm]
  \rcB \tf{occaVariable       } & \tf{                                                          } \wvs & \tf{            } & \tf{                } \\[1mm]
  \rcA \tf{occaRestrict       } & \tf{\_\_restrict\_\_                                          } \wvs & \tf{restrict    } & \tf{\_\_restrict\_\_} \\[1mm]
  \rcB \tf{occaVolatile       } & \tf{                                                          } \wvs & \tf{volatile    } & \tf{\_\_volatile\_\_} \\[1mm]
  \rcA \tf{occaConst          } & \tf{const                                                     } \wvs & \tf{const       } & \tf{const           } \\[1mm]
  \rcB \tf{occaAligned   } \wvs & \tf{\_\_attribute\_\_ ((aligned (\_\_BIGGEST\_ALIGNMENT\_\_)))} & \tf{                 } & \tf{                } \\[1mm]
  \hline
\end{tabular}
\captionof{table}{\occa keywords related to \occa variables attributes.}
\label{tab:memoryAttributes}
\end{center}\eMini\vspace{9mm}
}


{\scriptsize
\sMini\begin{center}\begin{tabular}{|l|l|l|l|}
  \hline
  \wvs OCCA                     & OpenMP                                                          & OpenCL                               & CUDA                                     \\ \hline
  \rcA \tf{occaKernelInfoArg   } & \tf{const int *occaDims } \wvs & \tf{\_\_global int *dims} & \tf{int *dims                } \\[1mm]
  \rcB \tf{occaFunctionInfoArg } & \tf{const int *occaDims,} \wvs & \tf{int \_dummy         } & \tf{int dummy                } \\[-1mm]
  \rcB \tf{                    } & \tf{int occaInnerId0,   } \wvs & \tf{                    } & \tf{                         } \\
  \rcB \tf{                    } & \tf{int occaInnerId1,   } \wvs & \tf{                    } & \tf{                         } \\
  \rcB \tf{                    } & \tf{int occaInnerId2    } \wvs & \tf{                    } & \tf{                         } \\[1mm]
  \rcA \tf{occaFunctionInfo    } & \tf{occaDims,           } \wvs & \tf{999                 } & \tf{1                        } \\[1mm]
  \rcA \tf{                    } & \tf{occaInnerId0,       } \wvs & \tf{                    } & \tf{                         } \\[1mm]
  \rcA \tf{                    } & \tf{occaInnerId1,       } \wvs & \tf{                    } & \tf{                         } \\[1mm]
  \rcA \tf{                    } & \tf{occaInnerId2        } \wvs & \tf{                    } & \tf{                         } \\[1mm]
  \rcB \tf{occaKernel          } & \tf{extern "C"          } \wvs & \tf{\_\_kernel          } & \tf{extern "C" \_\_global\_\_} \\[1mm]
  \rcA \tf{occaFunction        } & \tf{                    } \wvs & \tf{                    } & \tf{\_\_device\_\_           } \\[1mm]
  \rcB \tf{occaFunctionShared  } & \tf{                    } \wvs & \tf{\_\_local           } & \tf{                         } \\[1mm]
  \rcA \tf{occaInnerReturn     } & \tf{{continue;}         } \wvs & \tf{{return;}           } & \tf{{return;}                } \\[1mm]
  \hline
\end{tabular}
\captionof{table}{\occa keywords related to kernel prototypes and kernel setup.}
\label{tab:deviceKeywords}
\end{center}\eMini\vspace{9mm}
}


{\scriptsize
\sMini\begin{center}\begin{tabular}{|l|l|l|l|}
  \hline
  \wvs OCCA                     & OpenMP                                                          & OpenCL                               & CUDA                                     \\ \hline
  \rcA \tf{occaLocalMemFence  } & \tf{} \wvs & \tf{CLK\_LOCAL\_MEM\_FENCE } & \tf{                  } \\[1mm]
  \rcB \tf{occaGlobalMemFence } & \tf{} \wvs & \tf{CLK\_GLOBAL\_MEM\_FENCE} & \tf{                  } \\[1mm]
  \rcA \tf{occaBarrier(Fence) } & \tf{} \wvs & \tf{barrier(Fence)         } & \tf{\_\_syncthreads();} \\[1mm]
  \hline
\end{tabular}
\captionof{table}{\occa barriers needed in parallel threading synchronization.}
\label{tab:barriers}
\end{center}\eMini\vspace{9mm}
}


{\scriptsize
\sMini\begin{center}\begin{tabular}{|l|l|l|l|}
  \hline
  \wvs OCCA                     & OpenMP                                                          & OpenCL                               & CUDA                                     \\ \hline
  \rcA \tf{occaPrivateArray} & \tf{occaPrivateClass<type,sz> name} \wvs & \tf{type name[n]} & \tf{type name[n]} \\[1mm]
  \rcB \tf{occaPrivate     } & \tf{occaPrivateClass<type,1 > name} \wvs & \tf{type name   } & \tf{type name   } \\[1mm]
  \hline
\end{tabular}
\captionof{table}{\occa keywords expanding to platform-dependent private memory types and used to carry over loop-breaks in OpenMP due to barriers.}
\label{tab:privateVariable}
\end{center}\eMini\vspace{9mm}
}


{\scriptsize
\sMini\begin{center}\begin{tabular}{|l|l|l|l|}
  \hline
  \wvs OCCA                     & OpenMP                                                          & OpenCL                               & CUDA                                     \\ \hline
  \rcA \tf{occaCPU} & \tf{1} \wvs & \tf{0} & \tf{0} \\[1mm]
  \rcB \tf{occaGPU} & \tf{0} \wvs & \tf{1} & \tf{1} \\[1mm]
  \rcA \tf{occaOpenMP} & \tf{1} \wvs & \tf{0} & \tf{0} \\[1mm]
  \rcB \tf{occaOpenCL} & \tf{0} \wvs & \tf{1} & \tf{0} \\[1mm]
  \rcA \tf{occaCUDA}   & \tf{0} \wvs & \tf{0} & \tf{1} \\[1mm]
  \hline
\end{tabular}
\captionof{table}{\occa keywords specifying platform for platform-dependent kernel optimization.}
\label{tab:platformVariables}
\end{center}\eMini\vspace{9mm}
}


\bibliographystyle{plain}
\bibliography{references}

\end{document}